\documentclass{ws-procs9x6}

\begin{document}

\title{Gravitational Mediation of Supersymmetry Breaking in Superstring Theory}

\author{TOMASZ R. TAYLOR}

\address{Department of Physics,
Northeastern University\\
Boston, MA 02115 USA\\
E-mail: taylor@neu.edu}

\maketitle

\abstracts{SUSY breaking and its mediation are among the most important problems of supersymmetric generalizations of the standard model. The idea of gravity-mediated SUSY breaking, proposed in 1982 by Arnowitt, Chamseddine and Nath, and independently by Barbieri, Ferrara and Savoy, fits naturally into superstring theory, where it can be realized at both  classical as well as quantum levels. This talk is dedicated to Pran Nath on his 
65th birthday.}

\section{History}

In the summer of 1982, here at Northeastern University, Pran Nath, together with Richard
Arnowitt and Ali Chamseddine,\cite{pran} constructed the first SUGRA (supergravity) GUT (Grand Unified
Theory) with supersymmetry breaking in a ``hidden'' sector, communicated to the electro-weak
sector by gravitational interactions, so that
\begin{equation}
M_{E-W}\sim\frac{M_S^2}{M_{PLANCK}},
\end{equation}
with the supersymmetry mass scale $M_S\sim 10^{11-12}$GeV.

At the same time, on the other side of Atlantic Ocean,
Riccardo Barbieri, Sergio Ferrara and Carlos Savoy,\cite{sergio} constructed very similar models.

Today, 22 years later, gravitational mediation of SUSY breaking remains as one of the basic
model-building blocks ``beyond the standard model.'' This talk will celebrate Pran's
pioneering work by describing how his ground-breaking ideas are implemented in modern superstring
theory.

The 1982 hidden sector contained just one chiral supermultiplet, which was a very modest
proposal as compared to superstring theory in 2004; a generic superstring compactification
is full of:
\begin{itemize}
\item moduli: scalar fields whose vacuum expectation values (VEVs) determine the size and shape
of the compactified dimensions \item gauge fields, in bulk and on D-branes: non-perturbative
phenomena like gaugino condensation can lead to dynamical supersymmetry breaking in
strongly-interacting gauge sectors.
\end{itemize}
Moduli stabilization together with the related SUSY breaking and its mediation are among the most important problems of superstring phenomenology. Here are some (at least partial) solutions...

\section{SUSY Breaking and Moduli Stabilization}

Historically, there had been two routes to low-energy SUSY breaking: a field-theoretical
route and a string-theoretical one. The field-theoretical approach is based on a very
reasonable expectation that low-energy SUSY breaking is indeed a {\it low-energy\/}
phenomenon involving only the lowest (massless) superstring excitations, so that the heavy
superstring modes are just passive spectators. In this case, a single, well-understood
phenomenon like gaugino condensation could produce a non-trivial superpotential, stabilize
moduli VEVs and break SUSY at the same time. This mechanism works very well in the heterotic superstring theory, up to a small, but often
exaggerated problem of  a ``run-away'' dilaton VEV. In fact,
this run-away  occurs only when the hidden gauge
group is semi-simple; otherwise, individual gauge group factors create a ``racetrack'' that
generically locks the dilaton in a semi-stable vacuum which can be either supersymmetric or
non-supersymmetric, depending on the specifics of the model. The moduli
freeze in a stable configuration, which in orbifold models often corresponds to a special,
self-dual point in the moduli space. 
Moduli can also be stabilized in  Type II compactifications by switching on fluxes related to some higher-dimensional fields. In either case, if the vacuum is
non-supersymmetric, SUSY breaking is mediated by gravitational interactions
already at the classical level and the soft supersymmetry breaking terms are generated in exactly the same way as described in the original papers.\cite{pran,sergio}

Although phenomenologically viable, the
field-theoretical approach is not suitable for including quantum gravitational effects which
necessitate reintroducing string excitations just for the sole purpose of regulating
ultra-violet divergences. This can be mimicked by using
a good regularization scheme, for instance a Pauli-Villars procedure.\cite{maryk}
Here, we will approach this problem in a more direct way, in the framework
of ultra-violet finite string perturbation theory.

Any string theoretical mechanism that allows for an adjustable
supersymmetry breaking scale is in one way or another equivalent to
the Scherk-Schwarz (SS) breaking. It boils down to projecting out
string states that are odd under simultaneous $2\pi$ spacetime
rotation, i.e.\ $(-1)^F$ where $F$ is the fermion number, and a
translation by the full length $2\pi R$ of an extra-dimensional
circle, i.e.\ $(-1)^n$, where $n$ is the Kaluza-Klein momentum number. For this SS circle one can choose, for example, one of the
cycles of $T^2$ in $K3\times T^2$ compactifications. In this
way, one creates a mass gap in the fermion spectrum, with the
gravitino mass $m_{3/2}\sim 1/R$. All fermions from the closed
string sector acquire similar masses. However, if the theory
contains also an open string sector, with strings ending on
D-branes, the corresponding fermions acquire tree level masses only
if their D-brane world-volume encompass the SS direction. On the
other hand, if the D-brane is perpendicular to the SS direction, the
corresponding fermion spectrum is  not affected by supersymmetry
breaking, at least at the classical level.\cite{ads} The latter case looks
very interesting from both the phenomenological and theoretical
viewpoints because it offers a simple setup for studying quantum mediation
of supersymmetry breaking in  the calculable framework of
superstring theory.

\section{Bulk-To-Brane SUSY Breaking Mediation}

In Type I theory, open strings stretching between D-branes give rise
to gauge bosons and scalars representing D-brane motion in the
ambient space. After SUSY is broken in the bulk by SS mechanism,
these fermions acquire masses as a result of quantum loop effects.
In this talk, I will focus on gaugino mass generation.\cite{at}

When moving in spacetime, open superstrings sweep two-dimensional
world-sheet surfaces bordered by D-brane positions. In order to
couple to gauginos, at least one such a boundary is required.
The computation of a particular mass correction amounts to
integrating the two gaugino vertex operators over the boundary,
followed by an integral over the moduli of the {\em Riemann\/} surface
-- a formidable task indeed for a general world-sheet. The ``minimal''
surface is restricted by $U(1)$ charge conservation of the
underlying superconformal field theory:
\begin{equation}
g+\frac{h}{2}=\frac{3}{2},\end{equation} where $g$ is the number of
handles and $h$ is the number of boundaries. The solution with
$h=3$, $g=0$ represents a disk with two holes that can be
interpreted as a two-loop open string diagram. Since open strings do
not participate in SUSY breaking, this mass contribution is expected
to vanish, or more precisely, to be suppressed in the large $R$
limit as $e^{-R^2/\alpha'}$, where $\alpha'$ is the string scale. Hence we are left with $h=1$, $g=1$, a surface
with one boundary and one handle, shown on Figure 1.
 \begin{figure}[ht]
\centerline{\epsfxsize=2.1in\epsfbox{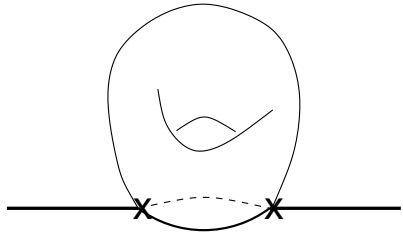}}
\caption{Genus 3/2 surface: two same-helicity gaugino vertex operators are inserted at the boundary. ~~~~\label{inter}}
\end{figure}
The  spectrum of
Kaluza-Klein excitations of closed strings propagating through the
handle is non-supersymmetric, therefore such a diagram can indeed
communicate SUSY breaking to open strings. We call it a ``genus
3/2'' surface, although more properly is should be called an
``orientable genus 1 surface with Euler characteristics $-1$.''
It can be thought of as a torus with a hole in its
surface. It is characterized by one complex modulus $\tau$ which controls the ``thickness'' of the handle and one real modulus $l$ which determines the size of the hole, or more precisely, the width of the throat between the boundary and the handle. According to the standard power counting rules for the string coupling constant $g$, this diagram comes at order $g$, and since the gaugino kinetic terms are of order $1/g$, the mass will be of order $g^2$.
Alernatively, one can replace the string coupling constant by the coupling constant $\lambda$ of the gauge group associated to the D-brane: then 
$g^2=\lambda^4$.

For a generic SS radius $R$, the computation of the gaugino mass is very difficult. In the case of $K3$ orbifold compactifications however, 
all steps can be made very explicit. One finds that a very important role is played by orbifold symmetries. As an example consider a $T^4/\mathbb{Z}_2$ orbifold, with $\mathbb{Z}_2$ acting as a simultaneous reflection
in the two complex planes of $T^4$. There is a residual symmetry with respect to interchanging these two complex planes. It turns out that the gaugino mass operator is odd under this symmetry, therefore the mass vanishes!
Furthermore, SS compactifications are also symmetric with respect to reflections
of the SS circle, with the momentum number $n\to -n$, and this symmetry
is realized in effective field theory as a discrete subgroup of $U(1)$ R-symmetry associated to the ``third'' complex plane. We conclude that one has to violate quite a few symmetries before generating a non-vanishing string loop correction the the gaugino mass. For instance, the orbifold symmetries can be eliminated by blowing-up the orbifold singularities
with VEVs of some twisted fields. This amounts to inserting one additional vertex in the bulk of the world-sheet.

Superstring computations can be simplified by focussing
on the large SS radius behavior of the mass.
The reason is that in the $R\to\infty$ limit, the spectrum of Kaluza-Klein states propagating through the handle becomes almost supersymmetric, with fermion
contributions canceling bosons unless they propagate for a 
proper time sufficiently long to allow the whole Kaluza-Klein tower to contribute. This corresponds to $\tau\to\infty$, a ``pinching'' 
limit of the handle in which our surface degenerates to a disk radiating and reabsorbing a massless closed string excitation, see Figure 2. 
\begin{figure}[ht]
\centerline{\epsfxsize=3.1in\epsfbox{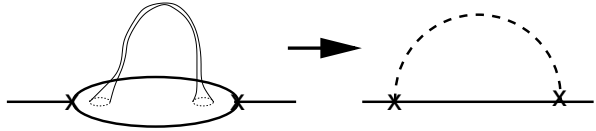}}
\caption{$\tau\to\infty$  limit of genus 3/2 and its further degeneration to a gravitational loop. ~~~~\label{interb}}
\end{figure}
As shown on Figure 2, by a further shrinking of the disk, that is by throwing away all except for the lightest ``virtual'' open strings, one can distort the world-sheet to a standard one-loop Feynman diagram representing a radiative gravitational correction!
The result of the full string computation is:
\begin{equation}
m_{1/2}\propto \lambda^4 m_{3/2}^3
\end{equation}
This can be explained in the following way.
The Feynman diagram shown on Figure 2 contains two powers of the gravitational coupling that bring the factor $M_{PLANCK}^{-2}$, and it is quadratically divergent
in the ultra-violet regime, so one expects
\begin{equation}
m_{1/2}\propto m_{3/2}\,\frac{\Lambda_{UV}^2}{M_{PLANCK}^{2}}
\end{equation}
where $\Lambda_{UV}$ is the ultra-violet cutoff. With supersymmetry
broken at $m_{3/2}$ one expects $\Lambda_{UV}\sim m_{3/2}$
and $m_{1/2}\propto m_{3/2}^3$. Indeed, one can reproduce Eq.(3) by using Pauli-Villars regularization.\cite{mkg} Hence the string loop correction can be understood as quantum-gravitational mediation  at the one-loop level.

The string derivation of Eq.(3) contains a very interesting subtlety which caused some confusion in the original version of Ref.[5].
In fact, there is another way of viewing Figure 1: in the limit $l\to 0$, the genus 3/2 surface splits into a torus and a disk connected by a narrow tube, as shown in Figure 3. 
\begin{figure}[ht]
\centerline{\epsfxsize=3.1in\epsfbox{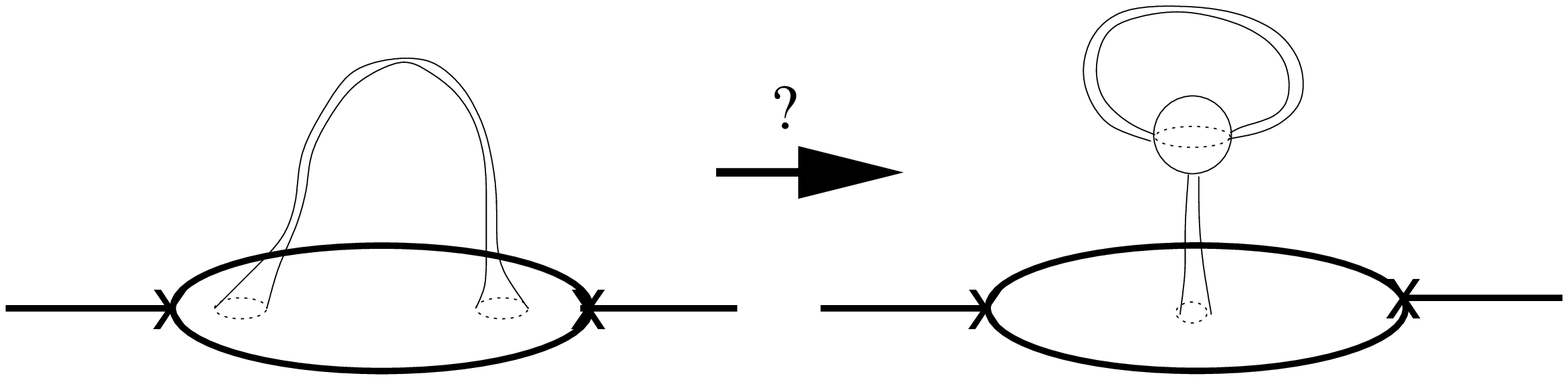}}
\caption{$l\to 0$  limit of genus 3/2. ~~~~\label{interc}}
\end{figure}
This region of moduli space cannot be interpreted as gravitational mediation.
In principle, it could yield contributions as large as $m_{1/2}\propto\lambda^4 m_{3/2}$. The reason why we do not find such an ``anomalously'' large mass is that there are no massless particles  propagating between the disk and the loop. This is related to the puzzling absence of anomaly mediation,\cite{amed}
which is believed to be a universal mass generation mechanism
in spontaneously broken SUGRA models.

Anomaly mediation is an important ingredient in many supergravity-based
generalizations of the standard model. There are several, equally confusing ways of explaining it, but the idea is that once SUSY is broken in the gravitational sector, the same loop corrections that give rise to superconformal anomaly, generate also some mass terms for gauginos.
It would be very nice to see this mechanism at work in a model with a sensible ultra-violet completion. In particular, superstrings
with Scherk-Schwarz SUSY breaking provide a natural testing ground for
these ideas. If one looks, however, more carefully at the ``anomaly-mediated'' formulas, one finds an unpleasant surprise that this mechanism does not yield a mass in no-scale supergravity models, which is exactly in the case
of SS breaking. 
It is quite disappointing
that anomaly mediation does not operate even at higher, $\lambda^4$ order in string loops.

\section{Conclusions}
22 years later, we are still struggling to understand SUSY breaking, now in superstring theory. The minimal SUGRA model will be soon tested at LHC. Just wait three or four more years to see the gravity-mediated spectrum of gauginos, squarks and sleptons... Congratulations, Pran, on your great ideas!

\section{About Pran}
When I came to CERN, on my first postdoc straight from Poland, where
a tough winter of 1981 was coming up, I noticed very interesting
preprints coming from Northeastern University in a faraway Boston.
They were co-authored by Arnowitt, Chamseddine and Nath, and there
were so many of them! It was quite intriguing that the order of
author's names kept changing in an unpredictable pattern. So
naturally, I started wondering who Pran Nath was. Little did I know
that few years later, I would come to Northeastern to join Pran as a
faculty collegue.

I have enormous respect for Pran. He truly deserves to be called a 
scholar: he
believes in physics as an important part of human civilization and
forcefully defends research in fundamental theory. He takes every
idea seriously and works very, very hard, pursuing his own dream of
a theory of everything. Happy Birthday, my friend! We are looking
forward to hundreds more of your papers.

\section*{Acknowledgments}
I am grateful to Ignatios Antoniadis for a fruitful collaboration.
I would like to thank Mary K. Gaillard for a very illuminating discussion and correspondence that clarified the field-theoretical interpretation of our results. This work was supported in part by the National Science Foundation under grant PHY-02-42834. Any opinions, findings, and conclusions or recommendations expressed in this material are those of the author 
and do not necessarily reflect the views of the National Science Foundation.

\section*{Note Added}
After this paper was submitted to the Pran Nath Festschrift, the discussion of Ref.[5] has been extended and improved in Ref.[8]. A new type of non-gravitational mediation of SUSY breaking between open string sectors has been discussed in Ref.[9].

\end{document}